\documentclass[11pt,a4paper]{article}

\usepackage{amssymb}
\usepackage{graphicx}

\newcommand{\be}{\begin{equation}}
\newcommand{\ee}{\end{equation}}
\newcommand{\bea}{\be \begin{array}{rcl}}
\newcommand{\eea}{\end{array}\ee}
\newcommand{\ra}{\rightarrow}

\newcommand{\av}[1]{\langle #1 \rangle}

\newcommand{\captionW}[1]{\parbox{12.5cm}{\caption{\footnotesize #1}} }

\newcommand{\half}{{\frac{1}{2}}}
\newcommand{\third}{{\frac{1}{3}}}

\newcommand{\dt}[1]{\partial_t #1}

\newcommand{\bra}[1]{\langle{#1}|}
\newcommand{\ket}[1]{|{#1}\rangle}
\newcommand{\hh}[2]{\bra{#1}\hat{H}\ket{#2}}

\newcommand{\cx}{c}
\newcommand{\cy}{c'}

\newcommand{\ha}{\hat{a}}
\newcommand{\had}{\hat{a}^{\dagger}}

\title{Particle Trajectories for Quantum Field Theory}
\author{Jeroen C. Vink  \thanks{
                               Shell Global Solutions International B.V.,
                               Kesslerpark 1, 2288GS Rijswijk, 
		                      The Netherlands.
							  Email: Jeroen.Vink@Shell.com} }

\begin{document}


\maketitle

\abstract{ The formulation of quantum mechanics developed by Bohm, which can generate 
well-defined trajectories for the underlying particles in the theory, can equally well be applied 
to relativistic Quantum Field Theories to generate dynamics for the underlying fields. 
However, it does not produce trajectories for
the particles associated with these fields. Bell has shown that an extension
of Bohm's approach can be used to provide dynamics for the fermionic
occupation numbers in a relativistic Quantum Field Theory.
In the present paper, Bell's formulation is adopted and elaborated on, with a full account of all technical
detail required to apply his approach to a bosonic quantum field theory on a lattice.
This allows an explicit computation of (stochastic) trajectories for massive and
massless  particles in this theory. Also particle creation and annihilation, and their impact
on particle propagation, is illustrated using this model.
}


\section{Introduction} \label{sect1}
Classical theories provide an intuitive and satisfactory way to explain the world, because 
they provide the dynamics for the ``elements of reality''. 
I.e., classical physics theories describe 
how particles move or how electromagnetic fields evolve as a function of time. This type 
of representation or explanation of the physical world is satisfactory, because it is 
fully self-contained and does not require any external (or internal) observers.
This essential property of a classical physics theory, implies that
it is possible (in principle) to write a
computer program that implements the concepts and dynamics of this  
theory, such that the ensuing numerical simulation of the system, 
even if this pertains to a (possibly dramatically) simplified version of the world,
will faithful represent the appropriate elements of this world along with their dynamics. 
I.e., at any point in the simulation, there will be values for internal 
variables, that directly translate to observations in the simulated world. 
For example, when classical Maxwell theory is simulated, where particles interact 
through electromagnetic forces, the simulation will produce 
the time evolution of the particle positions and electromagnetic field configurations.

Providing such a ``beable'' \cite{Bell}, or ``simulatable'' representation of the world
poses a problem for quantum theories. Quantum mechanics in its 
standard formulation describes the dynamics of the Schr\"odinger wave function, and 
this time evolution can of course be simulated on a computer. However, it is then not 
clear how to extract (compute or display) the expected observable properties of 
the beables in such a simulation: In contrast to classical theories, actual particle locations 
and field values are not directly represented in standard formulations of quantum theories and 
one needs to introduce an external observer, which cannot be included in the dynamics of the 
system, to make the link with desired concepts like particle positions or field values.

The only interpretation, or rather formulation of (non-relativistic) quantum mechanics 
that achieves the same level of realism as classical mechanics, is the formulation originally 
proposed by de Broglie \cite{deBroglie56}, which was rediscovered and fully developed 
by Bohm \cite{Bohm52,BohmHiley}. Here, as in a classical system, particle locations are 
part of the theory and have a (causal) dynamics that can be computed from the dynamics 
of the wave function of the system. In this way the measurement problem alluded to above 
is solved, and the theory can be used to represent 
physical systems of any level of complexity in a fully self-contained manner. Hence, 
using Bohm's approach, we can simulate movement of a single particle that tunnels through 
a potential barrier \cite{VinkNVVN}, or study the quantum dynamics of the universe as a whole,
even when it is represented using only a few degrees of freedom, as in 
Wheeler-DeWitt ``mini-super space'' \cite{Vink92,GlikmanVink90}. 

It may be noted in passing, that also Everett's many worlds interpretation of quantum
mechanics \cite{Everett57}  fails this ``simulatability test''. For the beables to materialize in the 
many worlds formulations, one first needs to introduce an external and arbitrary mechanism to 
trigger a splitting of the wave function into its many possible beable eigenstates - 
for simple quantum systems one cannot use some kind of ``level of decoherence'' as a trigger.
Second, if we accept an arbitrary splitting-trigger as a extra ingredient in the system's dynamics, the resulting 
exploding number of worlds, would have beable values that are 
random samples from the evolving probability distribution 
defined by the wave function. Hence, it would literally be impossible to identify or keep track of
a single world in such a computer simulation, in which the particles follow recognizable trajectories.

Bohm's formulation of quantum mechanics appears to be gaining acceptance, for example 
in the quantum chemistry community \cite{QuantChem}. Here it is used to provide 
alternative ways to inspect or probe the quantum systems. Also, the alternative 
formulation might lead to possibly more efficient methods to perform computer 
simulations of quantum systems \cite{Poirier10,Vink2004}.
However, Bohm's formulation has not gained general acceptance (yet) and the Copenhagen 
interpretation remains to be favored by the majority of physicists. The main criticism, 
or reason for distrust of Bohm's formulation is the apparent difficulty to extent the 
formulation to relativistic Quantum Field Theory (QFT) \cite{Bell,Struyve11}. 

This mistrust, however, is misplaced, as has been argued elsewhere \cite{Struyve11,DurrGoldsteinZanghi,DurrGoldstein}
and will be explicitly demonstrated also in this paper.
The immediate, and most obvious extension to quantum field theory, was already formulated 
by Bohm in his early 1952 papers \cite{Bohm52}. 
Here he used the Hamiltonian formulation of QFT, in which a (functional) Schr\"odinger 
equation governs the dynamics of a wave functional that has the field as argument. 
Bohm showed that the same approach to find particle trajectories from a wave function
that has particle location as argument, can be applied to a wave functional 
that has a field as argument. Hence, one can obtain causal dynamics for the fields - 
in Bohm's example he obtained dynamics for the electromagnetic fields.
This is gratifying, since we can now perform a computer simulation of a (properly regularized)
pure QED (i.e., electromagnetics without electrons or other particles) that provides 
the dynamics of the electromagnetic fields (or vector potential) - just like we
can do for the classical Maxwell theory. However, Bohm's formulation is not 
sufficient, since  simulations using his formulation cannot produce trajectories for the
photons that should also be present in this theory; nor can it produce trajectories for
the electrons and positrons of full QED.
Hence, when applied to QFT, Bohm's formulation can not resolve the particle-wave duality that was so 
compellingly explained by his formulation of non-relativistic quantum mechanics.

This is where Bell steps in. In his beautiful paper ``Beables for Quantum Field Theory" 
\cite{Bell}, he explains how a generalized version of Bohm's formulation can be 
applied to QFT to provide dynamics for particle location. More specifically, he describes
how dynamics for fermionic occupation numbers can be obtained from a 
lattice-regularized QFT. Unfortunately, his description is equally succinct as it is profound 
and it is not immediately clear what the nature of the particle trajectories will be,
when his prescription is applied to an 
actual QFT. For example, Bell's proposed dynamics is stochastic and it is not clear
to what extent this randomness would survive in the continuum limit of the 
lattice-regularized QFT. Also, Bell only considers fermions; the extension to bosons
appears to be straightforward, but it is not at all obvious what ensuing
trajectories for massless bosons, like photons, would look like.

A further investigation of the nature of Bell's approach
was done in ref.~\cite{Vink93} for non-relativistic quantum mechanics. 
For relativistic QFT, an application of Bell's approach was pursued in 
ref.~\cite{DurrGoldsteinZanghi} (see also \cite{DurrGoldstein}). Even though this 
paper provides much more detail, it deviates in its approach from Bell's more rigorous 
formulation in which a lattice-regularized QFT with well defined occupation 
numbers for the particles was assumed. In the version of D\"urr et al., a more hybrid description is
proposed, in which Bell's stochastic jumps in observable values are only adopted to 
explain particle creation and annihilation - in between the random branching and
merging of trajectories the particle dynamics 
follows from Bohm's (non-relativistic) causal prescription. Furthermore,
their formulation appears to require a split of the Hamiltonian in a free and 
interaction part - similarly to what is required in perturbative treatments to QFT. 
Such a split becomes problematic when non-perturbative phenomena,
such as quark confinement, have to be addressed.

This paper addresses the task to supply more detail to the description laid out in
ref.~\cite{Bell} and aims to show explicitly how particle trajectories 
are obtained by rigorously applying Bell's formulation to lattice QFT. Using lattice
QFT will (in principle) allow for numerical simulations of particle and field 
trajectories in the full Standard Model, addressing both perturbative and
non-perturbative phenomena.
The present paper will, of course, only take a small step in that direction.

The remainder of this paper is organized as follows. 
In sect.~\ref{Sect2}, we lay down the basic formulation of the 1+1 dimensional QFT that
is used as testing ground, and review in detail how Bell's proposal can be implemented 
in this theory. The resulting stochastic particle trajectories will be referred to as
``de Broglie-Bohm-Bell'' (BBB) trajectories throughout the paper.
In sect.~\ref{Sect3} we
provide numerical and (partial) analytical examples to explore the nature and characteristics
of the particle trajectories generated from this QFT. These trajectories are shown to
have the correct non-relativistic limit; for massive bosons the stochastic nature of the
trajectories is shown to be suppressed at sufficiently large scales and, perhaps most
surprisingly, massless bosons are found to have an average velocity equal to the speed
of light - but typically will continue to exhibit random jumps while propagating.
In sect.~\ref{Sect4} we show trajectories for two free bosons on a collision course and
in sect.~\ref{Sect5} we switch on interactions and show that spontaneous particle creation
naturally occurs when the self-interaction is sufficiently strong. This then leads to an
increase of the effective mass of the propagating particle.
Finally, in sect.~\ref{Sect6}, we summarize our findings and offer concluding remarks.

\section{Groundwork for BBB Trajectories in Quantum Field Theory} \label{Sect2}
In order to show how Bell's formulation can be applied to QFT, we shall use a theory 
that is simple enough to allow for (partial) analytical treatment 
as well as numerical simulations. 
Hence, we shall use a 1+1 dimensional theory for a scalar field with cubic self-interaction. 
Following Bell, and to ensure that all operations are well-defined, we use the 
lattice formulation of this theory. Details on lattice QFT can be found, for example, 
in refs.~\cite{Smit,Munster}; in particular we will follow the notations of ref.~\cite{Smit}.

The simplified QFT used in the following is defined on a 1-dimensional spatial and periodic lattice, 
with  coordinates $x$ given by $x=na$, $n = -N/2+1,\cdots,N/2$. The lattice 
distance is denoted by $a$, from which the size of the system follows as $L = Na$. 
To keep notations simple, we shall adopt natural units in which the velocity of 
light $c=1$ and Planck's constant $\hbar=1$; however, since it will be important
to keep track of lattice artifacts and how to approach the continuum limit, 
the lattice distance $a$, as well as the number of lattice sites $N$
(or equivalently, the size of the system, $L$) will be written explicitly. 
In the absence of interactions the continuum limit, $a \rightarrow 0$,
will typically be taken by increasing $N$ at a fixed value of $L$, such that $a/L = 1/N \rightarrow 0$. 
Following Bell, time is assumed to be continuous; however, in order to perform 
computer simulations also time will have to be discretized. Also for this time discretization 
it is useful to adopt methods from lattice QFT, to ensure for example that the time
evolution of the system's wave function is unitary.

In the Schr\"odinger picture, the Hamiltonian dynamics of a wave function 
can be written as,
\be
   i\partial_t \ket{\Psi} = \hat{H}\ket{\Psi}, \label{SchrEq}
\ee
where $\ket{\Psi}$ is the state vector in the system's Hilbert space and 
$\hat{H}$ is the Hamiltonian operator in this Hilbert space. 
For a scalar field with cubic self-interaction the Hamiltonian operator can be written as,
\be
 \hat{H} = \hat{H}_0 + \hat{H}_{int}, \label{Htot}
\ee
where the free particle Hamiltonian is
\be
 \hat{H}_0 = \half \sum_x \big( \hat{\pi}_x^2 + a^{-2}(\hat{\phi}_{x+a}-\hat{\phi}_x)^2 
                                         + \mu^2 \hat{\phi}_x^2 \big), \label{Hfree}
\ee
and
\be
  \hat{H}_{int}  =  \third \lambda \sum_x \hat{\phi}_x^3   \label{Hint}
 \ee
The mass parameter is denoted by $\mu$, the self interaction strength is $\lambda$, and the
summation is defined as
\be
\sum_x f(x) \equiv \sum_{n=-N/2 + 1}^{N/2} a f(an),
\ee
where the periodic boundary conditions imply that $f(x+L)\equiv f(x)$.
The somewhat unusual choice of a cubic self-interaction, is made to allow for transitions 
between one and two particle states, while keeping the numerical treatment to be discussed 
in section \ref{Sect5} below tractable. 

The Schr\"odinger equation for free fields, $\hat{H} = \hat{H}_0$, describes $N$ coupled harmonic
oscillators. This is perhaps most clearly seen when (\ref{SchrEq}) is expressed
in the field representation, with $\Psi(\phi) = \langle \phi\ket{\Psi}$, such that
the Schr\"odinger equation turns into
\be
   i\partial_t \Psi(\phi) = \frac{1}{2a^2} \sum_x \big( -\frac{\partial^2 \Psi(\phi)}{\partial \phi_x^2}
                   + \big( (\phi_{x+a}-\phi_x)^2 + (a\mu)^2\phi_x^2 \big)\Psi(\phi) \big).
																		\label{FieldRep}
\ee

The field operators $\hat{\phi}_x$ and field momentum operators $\hat{\pi}_x$ have 
canonical commutation relations,
\be
  [ \hat{\phi}_x,\hat{\phi}_y ] = [\hat{\pi}_x,\hat{\pi}_y] = 0; \;\; 
  [ \hat{\phi}_x,\hat{\pi}_y ] = i\delta_{x,y}.
\ee
The Kronecker delta with arguments $x=ma$ and $y=na$ is defined as
\be
   \delta_{x,y} = a^{-1}\delta_{m,n}.
\ee
The scalar field and its conjugate momentum can be written in terms of creation 
and annihilation operators as,
\bea
 \hat{\phi}_x & = & \sum_p (2 \omega_p)^{-1/2}\big( \ha_p e^{ipx} + \had_p e^{-ipx} \big), \\
 \hat{\pi}_x  & = & -i\sum_p (\omega_p/2)^{1/2}\big( \ha_p e^{ipx} - \had_p e^{-ipx} \big), \\
\eea
where the summation is over lattice momenta $p=2\pi k /L$,
\be
 \sum_p f(p) \equiv \frac{1}{L}\sum_{k=-N/2 + 1}^{N/2} f(2\pi k/L),
\ee
and the energies of the momentum eigenstates are
\be
  \omega_p = \big(\mu^2 + a^{-2}(2 - 2\cos(ap))\big)^{1/2}. 
\ee
The free field Hamiltonian can then be written as,
\be
  \hat{H}_0 = \half \sum_p  \omega_p (\had_p\ha_p + \ha_p\had_p).   \label{Haa}
\ee
The creation and annihilation operators obey the canonical commutation relations,
\be
 [\ha_p,\ha_q] = [\had_p,\had_q]=0; \;\; [\ha_p,\had_q]=\delta_{p,q}.
\ee
The Kronecker delta with lattice momenta $p=2\pi k/L$ and $q=2\pi l/L$,  is defined as
\be
   \delta_{p,q} = L \delta_{k,l}.
\ee
The creation operators can be used to generate multi-particle states from the vacuum 
$\ket{0}$ (which is defined by $\ha_p\ket{0} = 0$). 

All the above is standard for the Hamiltonian formulation of QFT for a real scalar 
field on a 1-dimensional spatial lattice. To make contact with the formulation of 
Bell \cite{Bell}, we follow ref.~\cite{DurrGoldsteinZanghi} and define
the creation operator for a particle at location $x$ as,
\be
   \had_x = \sum_p \had_p e^{ipx}. \label{ax}
\ee
These operators can be used to create multi-particle states in which the particles
have fixed locations, rather than fixed momenta. We shall use Fock basis states that explicitly
show the total number of particles $M$ in each state,
\be
   \ket{x_1\cdots x_M} = {\cal N}_M \had_{x_1}\had_{x_2}\cdots \had_{x_M} \ket{0}.
       \label{ketx}
\ee
Since the creation operators commute, i.e., the bosonic particles are indistinguishable,
we label the position states using a fixed order for the particle locations, 
such that $ x_1 \le x_2 \cdots \le x_M$. 
Usually the occupation number $n_i$ at lattice site $i$ will be either zero or one, but since the particles are
bosons, there could be multiple particles at the same location; in that case there will be multiple
occurrences of this lattice location in the state (\ref{ketx}). 
Finally, the constant
${\cal N}_M = (\prod_i n_i !)^{-1/2}$ normalizes the states, to ensure that
\be
    \bra{x_1\cdots x_M} y_1\cdots y_{M'}\rangle = 
	                          \delta_{M,M'}\delta_{x_1,y_1}\cdots \delta_{x_M,y_M}.
\ee 

We shall furthermore
assume that these multi-particle states span the full, physical, state spaces. I.e., we will
assume that the resolution of unity,
\be
 1 = \ket{0}\bra{0} + \sum_{x_1}\ket{x_1}\bra{x_1} + 
                 \sum_{M=2}^{M_{max}} \sum_{x_1\le \cdots \le x_M}
                   \ket{x_1 \cdots x_M} \bra{x_1 \cdots x_M}. \label{unity}
\ee
holds, provided that the cut-off value $M_{max}$ for the maximum number of particles is 
sufficiently large.  This is an important assumption, since it will allow switching between a 
field-representation of the quantum dynamics, as shown in Eq.~(\ref{FieldRep}), to a 
particle based representation, which will be explored further below. 
Fortunately, it is quite reasonable to assume that the
identity (\ref{unity}) holds, since the configuration states (\ref{ketx}) span the same Fock
space as the momentum states generated by acting with the momentum-type creation operators
on the vacuum. This Fock space is generally assumed to be
sufficiently large, to not only express the perturbative physics of a QFT, 
but also its non-perturbative content. 

In order to apply Bell's prescription to obtain trajectories for particles, we need to identify an
operator with eigenvalues that can represent the ``beable'' particle locations.
Such a ``particle configuration'' operator can be defined using the configuration states 
(\ref{ketx}) as
\be
   \hat{C} = \sum_{M=1}^{M_{max}} \sum_{x_1\le \cdots \le x_M}
                   \ket{x_1 \cdots x_M} c_x^{(M)} \bra{x_1 \cdots x_M}. \label{config}
\ee
It is clear that the state (\ref{ketx}) is an eigenstate of this operator with
eigenvalue $c_x^{(M)}$. This ``configuration index'' is an integer number
that must uniquely identify the locations $x_1,\dots,x_M$ of the particles in the corresponding
configuration eigenstate $\ket{x_1 \cdots x_M}$. It can be defined, for example, as
\be
   c_x^{(M)} = \sum_{m=0}^{M-1} N^m + \sum_{i=1}^M (x_i/a)N^{i-1}. \label{cIndex}
\ee
Note that, using this definition, the index will give different values to configurations 
that differ by a  permutation.
Hence, when $m>1$ only a subset of the $N^m$ index values will be used to label the physically different
configurations of the $m$ particles in this $m$-particle sector of the model. 
The advantage of using this indexing is that it is not only easy to
compute $c_x^{(M)}$ from any set of particle locations $x_1,\dots,x_M$, but also
straightforward for any value of $c_x^{(M)}$ to back-compute the number of
particles $M$, as well as the locations of these particles. Obviously, it is equally possible
to use a more compact, and numerically more efficient definition, which only labels the physically
distinguishable configurations.

Below, the time-dependent values of the configuration indices $c \equiv c_x^{(M)}$ will be used to  
represent the stochastic time evolution of particle locations. The eigenstates of
this configuration operator (\ref{config}) may be written as $\ket{\cx}$. 
Since $\cx$ is dimensionless, the completeness and orthonormality
conditions can then be written simply as
\be
   \sum_{\cx} \ket{\cx}\bra{\cx} = 1, \;\; \bra{\cx}\cy\rangle = \delta_{\cx,\cy}.
\ee
Since different value for the configuration index $\cx$
may represent different numbers of particles, the evolution of this index will not only
express particle trajectories, but also particle creation and annihilation.

Using the operator (\ref{config}), we can now follow the steps outlined in ref.~\cite{Bell}, 
to find the dynamics for the configuration index $\cx$, from which the locations of the particles 
in the configuration can be back-computed using Eq.~(\ref{cIndex}).
Given that the quantum state $\ket{\Psi}$ for the scalar field evolves according to 
the Schr\"odinger equation (\ref{SchrEq}), it follows that the  probability for the particle 
configuration to transition from $\cx$ to $\cy$ in the time interval $\delta t$ can be written as 
\be
   T_{\cy,\cx} \delta t = {\rm max}(0, J_{\cy,\cx} )  / P_{\cx}. \label{transA}
\ee
The $J_{\cy,\cx}$ combines the (time independent) matrix elements of the Hamiltonian in the 
configuration representation with specific values of the wave function in the configuration 
representation as,
\be
  J_{\cy,\cx} = 2{\rm Im}\big( \langle{\Psi}\ket{\cy} 
                         \bra{\cy} \hat{H} \ket{\cx} \langle{\cx}\ket{\Psi} \big).
\ee
$P_{\cx}$ in Eq.~(\ref{transA}) is the probability for the configuration labeled by index $\cx$
and is defined as
\be
 P_{\cx} = |\langle{\cx}\ket{\Psi}|^2.
\ee

It will be convenient to express the wave functions in the $\cx$-representation, such that
\be
   \ket{\Psi}=\sum_{\cx}\psi_{\cx}\ket{\cx}, \label{Crep}
\ee
where the coefficients $\psi_{\cx} = \langle \cx \ket{\Psi}$ are now time-dependent 
functions of the configuration index $\cx$, or equivalently, of the (discretized) locations 
$x_1,\dots,x_M$. 
Hence, the coefficients in (\ref{Crep}) can also be written as $\psi(x_1,\dots, x_M)$, which 
makes them resemble $M$-particle wave functions in which the particles move on 
a 1-D periodic lattice.
The Schr\"odinger equation (\ref{SchrEq}) in this particle configuration representation
turns into
\be
   i\dt \psi_{\cx} = \sum_{\cy}\bra{\cx}\hat{H}\ket{\cy} \psi_{\cy}.
										\label{SchrEqOneP}
\ee
or equivalently,
\be
   i\dt \psi(x_1,\dots,x_M) = 
        \sum_{y_1\le y_2\dots\le y_{M'}}\bra{x_1\dots x_M}\hat{H}\ket{y_1 \dots y_{M'}} 
                                                             \psi(y_1,\dots,y_{M'}).
										\label{SchrEqOnePx}
\ee

Similarly, in the $\cx$-representation, the transition probability rates can be written as
\be
   T_{\cy,\cx} = {\rm max}\big(0, 2{\rm Im}( \psi^*_{\cy} 
                 \bra{\cy} \hat{H} \ket{\cx} \psi_{\cx}) \big) /|\psi_{\cx}|^2.
                                                 				 \label{trans}
\ee

As is clear, for example, from the representation in creation and annihilation
operators (\ref{Haa}), the free-field Hamiltonian $H_0$ only has matrix 
elements between states with equal numbers of particles. This implies that a 
wave function that initially describes an $M$-particle state -- i.e., is a linear 
combination of $M$-particle eigenstates $\cx$ -- will remain an $M$-particles state 
also for later times. Obviously (but reassuringly), this carries over to the
transition probabilities (\ref{trans}) and there cannot be transitions 
between configurations with different numbers of particles. Hence, in the free theory, 
it is possible to self-consistently explore the trajectories of a single scalar particle - 
which we shall do in the next section.

\section{BBB Trajectories for a Free Scalar Boson} \label{Sect3}
In this section we shall focus on free QFT, where the interaction term $H_{int}$
in the Hamiltonian (\ref{Htot}) is zero. 
As was announced already above, we shall first focus on the $1$-particle sector. 
In this case the configuration
index is equivalent to the coordinate of the particle, $\cx\equiv x/a$. 
The $N$ orthonormal basis states that span this sector of the Hilbert space can be written as
\be
  \ket{x} = \had_x\ket{0},\;\;(x=-L/2+a,\dots,L/2),
\ee
and the corresponding matrix elements of the Hamiltonian are
\be
   \bra{y} \hat{H} \ket{x} =  \bra{0}\ha_y \sum_p \omega_p 
                                        (\had_p\ha_p + \half L) \had_x\ket{0}.
		\label{Hxy}
\ee
Using the mixed commutation relations,
\be
   [\ha_x,\ha_p]=[\had_x,\had_p] = 0, \;\; [\ha_x,\had_p] = e^{ipx},
                                               \;\;[\ha_p,\had_x] = e^{-ipx},
\ee
it follows that
\be
  \bra{y} \hat{H} \ket{x} = \sum_p \omega_p e^{ip(y-x)} + E_0 \delta_{y,x}
                                                                       \label{HxyOneP}
\ee
with $E_0 = \half L\sum_p \omega_p$ the vacuum energy contribution.
The transition matrix, limited to the $1$-particle sector, turns into
\be
  T^{(1)}_{y,x} = {\rm max}\big(0, {\rm Im}( 2\sum_p \omega_p e^{ip(y-x)}\psi^*(y)/\psi^*(x)) \big)
                 + E_0\delta_{y,x}. \label{TransOneP}
\ee
The vacuum energy term $\propto \delta_{y,x}$ can be absorbed in the overall normalization of 
$T^{(1)}_{x,x}$, which must be such that 
\be
aT^{(1)}_{x,x}\delta t = 1 - \sum_{y \ne x} T^{(1)}_{y,x}\delta t. \label{normT}
\ee

\subsection{Non-relativistic Limit} \label{SubSect3dot1}
As a first step in exploring the trajectories generated with the transition
rates of Eq.~(\ref{TransOneP}), it is worthwhile to investigate the 
non-relativistic limit, which can be achieved by letting $\mu \rightarrow \infty$. 
The mass dependence of the transition probabilities is contained in the
$\omega_p$-terms in the Hamiltonian. Expanding these terms in powers of $1/a\mu$ gives,
\bea
   \omega_p & = & \big( \mu^2 + a^{-2}(2 - 2\cos(ap)) \big)^{1/2} 
              = \mu \big(1 +(1 - \cos(ap))/(a\mu)^2 \big)^{1/2} \\
            & = & \mu + (1 - \cos(ap))/a^2\mu  + \cdots
\eea
For $a \ra 0$, this expansion reduces to an expansion in powers of $p^2/\mu^2$, which is 
appropriate for exploring  non-relativistic behavior.

The Hamiltonian in the 1-particle sector then follows from (\ref{HxyOneP}) as,
\be
   H_{y,x} = 
      a^{-1}\sum_p \left( e^{ip(y-x)}\big(a\mu + (1 - \cos(ap))/a\mu 
	                          + \cdots \big) \right) + E_0 \delta_{y,x}. \label{HwithM}
\ee
Writing the cosine as a sum of exponentials and using the identity,
\be
   \sum_p e^{-ip(y-x)} = \delta_{y,x},
\ee
the summation over momenta can be done for each term in the expansion, leading to 
\be
   H_{y,x} = (E_0 + \mu)\delta_{y,x} + 
          (2\delta_{y,x} - \delta_{y,x+a} - \delta_{y,x-a})/2a^2\mu + \cdots.
		  \label{H_xy}
\ee
The leading term in this $p^2/\mu^2$ expansion can again be absorbed in the vacuum energy. 
If then all but the next-leading terms are ignored, 
the Hamiltonian is exactly the same as the one obtained in quantum mechanics
for a free particle moving on a discretized circle (cf. \cite{Vink93}). 

The transition rate for the particle to move from location $x$ to $y$ ($y \ne x$) 
follows easily from (\ref{Hxy}) and (\ref{TransOneP}) as,
\be
   T^{(1)}_{y,x} = {\rm max}\left( 0, {\rm Im}
         \big((\delta_{y,x+a} + \delta_{y,x-a})/a^2\mu + \cdots \big)\psi(y)/\psi(x) \right).
             \label{ToneP}
\ee
As before, only the off-diagonal elements of $T$ are important.
The probability for a particle to stay at the same location is determined from the
the overall normalization of transition probabilities, as shown in Eq.~(\ref{normT}). 
The  complex conjugation of the wave function ratio has been replaced 
by a sign-flip of the imaginary part in (\ref{ToneP}).

We can now use the transition rates (\ref{ToneP}) to compute the expected value 
of the particle velocity; i.e., the average displacement of this particle in one time step $\delta t$.
To do this easily, we use a simple plane wave solution of the 1-particle Schr\"odinger equation 
(\ref{SchrEqOneP}): 
\be
   \psi_0(x) = L^{-1/2}e^{ip_0 x}. \label{psix}
\ee
Even though this now exactly looks like a plane wave solution for a particle on a 1-D lattice 
in normal quantum mechanics, it of course still represents a solution of the full QFT
Schr\"odinger equation (\ref{SchrEq}). This state represents a Fock state with one particle
that has a momentum $p_0$, and is an eigenstate of the free Hamiltonian with energy $\omega_{p_0}$.

The average velocity of the particles moving according to the transition rates (\ref{ToneP}) 
can be computed as
\bea
   \av{v} & = & \sum_y T^{(1)}_{y,x} (y - x) \\
          & = & \sum_y {\rm max}\left( 0, {\rm Im}
         \big( (\delta_{y,x+a} + \delta_{y,x-a})/a^2\mu + \cdots \big)e^{ip_0(y-x)} \right)(y - x) \\
		  & = & \sin( ap_0)/a\mu.    \label{NonRelLimit}
\eea
In the continuum limit, $a \ra 0$, this gives the expected result that the
(average) velocity is constant with a value of $p_0/\mu$. Since the trajectories are stochastic, it is
also important to check that the scatter around these straight-line trajectories is unobservably small
for macroscopic particle movements. This was done in ref.~\cite{Vink93} and will be revisited below. 

This result for the trajectories of low-energy particles is reassuring, but should not come as a
surprise, since it is well known that scalar QFT reduces to standard quantum mechanics in the
non-relativistic limit. In fact, it is easy to show in the free theory that Bohm's prescription to obtain particle trajectories
follows from the non-relativistic transition probabilities (\ref{ToneP}) in the continuum limit. 
The average velocity for a system with an arbitrary (1-particle) wave function $\psi(x)$ follows as
(cf.\ \ref{NonRelLimit})
\bea
   \av{v} & = & \sum_y {\rm max}\left( 0, {\rm Im} 
      \big( (\delta_{y,x+a} + \delta_{y,x-a})/a^2\mu \big)\psi(y)/\psi(x) \right)(y - x) \\
		 & = & {\rm Im} (\psi'(x)/\psi(x) ) = S'(x)/\mu. \label{CausalTraj}
\eea
Here we used $\psi(x+a) \approx \psi(x) + a\psi'(x)$ and wrote $\psi(x) = |\psi(x)|e^{iS(x)}$.
If the limit $a\ra 0$ is taken such that the velocity fluctuations vanish and $\av{v} = v$, we recover
Bohm's familiar prescription for the velocity $v$ of a particle associated with a wave function $\psi$.

In the next two subsections, we shall investigate the trajectories of massive and massless
relativistic particles respectively. For particles with finite or zero mass, the
transition probabilities are no longer ultra-local (as in Eq.~(\ref{ToneP}), where only jumps to
nearest neighbor lattice locations are allowed). Hence, it is not clear if the
stochastic nature of the trajectories will disappear in the continuum limit.

\begin{figure}[tbh] 
\begin{center}
\includegraphics[scale=0.41]{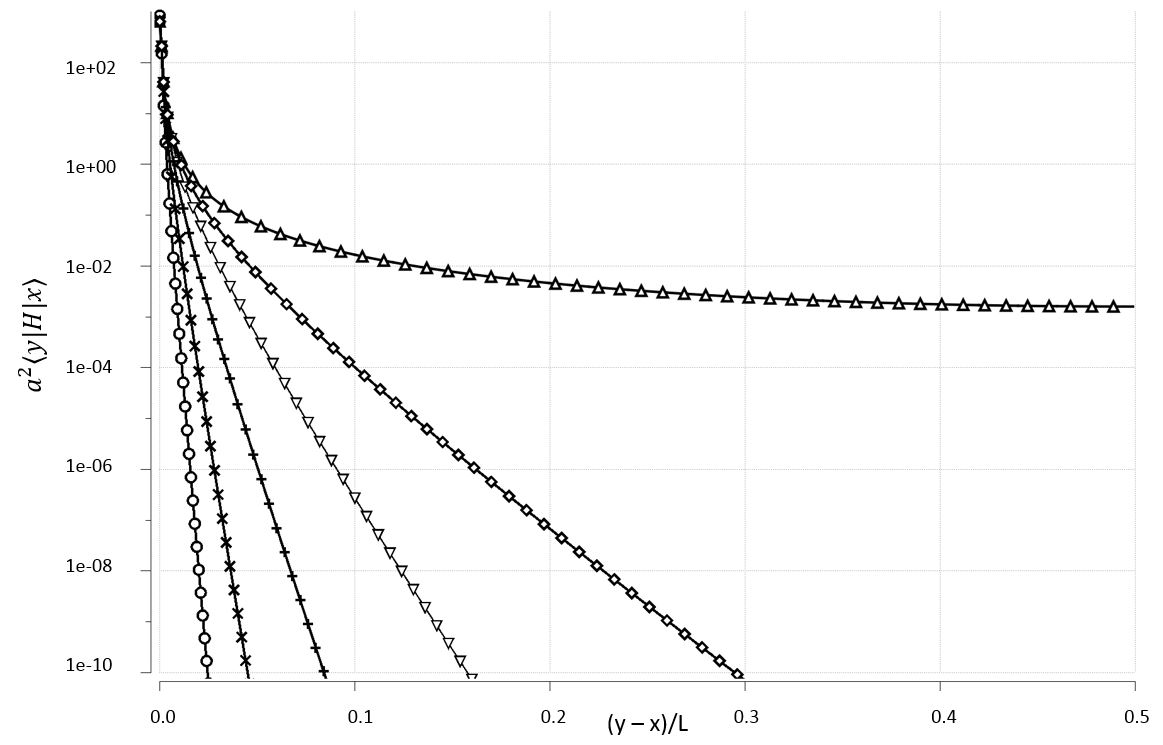}
\captionW{Locality of the Hamiltonian matrix elements. The curves show $\bra{y}\hat{H}\ket{x} $ 
on a logarithmic scale as a function or $(y-x)/L$, for $a\mu=0\, (\triangle), 0.0625\, (\diamond), 
0.125\, (\triangledown),  0.25\, (+), 0.5\,(\times)$ and $1\, (\circ)$ on a lattice with $1000$ sites.     \label{H_Kernel}}
\end{center}
\end{figure}

\subsection{Trajectories for Massive Particles} \label{SubSect3dot2}
When the particles have a finite (but non-zero) mass, the momentum sum in the Hamiltonian
matrix elements (\ref{HwithM}) cannot be evaluated analytically. However, it is straightforward
to perform the summation numerically. In this way the $x$ and $y$ dependence of the Hamiltonian 
matrix elements can be exposed. Since the system is translationally invariant, it is sufficient to
compute the matrix elements as a function of $|y-x|$. For $\mu \ra \infty$, only the nearest neighbor
elements, $|y-x| = a$, are non-zero. For finite mass, transitions to arbitrary
distant locations will be allowed; however, the probability should decrease (exponentially) with
distance. For dimensional reasons, one expects the decay-length (or correlation length) to
be $\propto 1/\mu$; i.e., equal to the Compton wave length of the particle. 
This behavior of $\bra{y}\hat{H}\ket{x}$ is confirmed by the results shown in 
Fig.~\ref{H_Kernel}. This plot shows the (logarithm) of the matrix elements as a function of $(y-x)/L$,
for a number of different values of the particle mass.
One recognizes an exponential decrease with distance - at least for intermediate distances that are
not affected by finite lattice size or finite volume effects. Moreover, the correlations length
indeed increases with decreasing particle mass.

\begin{figure}[bth] 
\begin{center}
\includegraphics[scale=0.37]{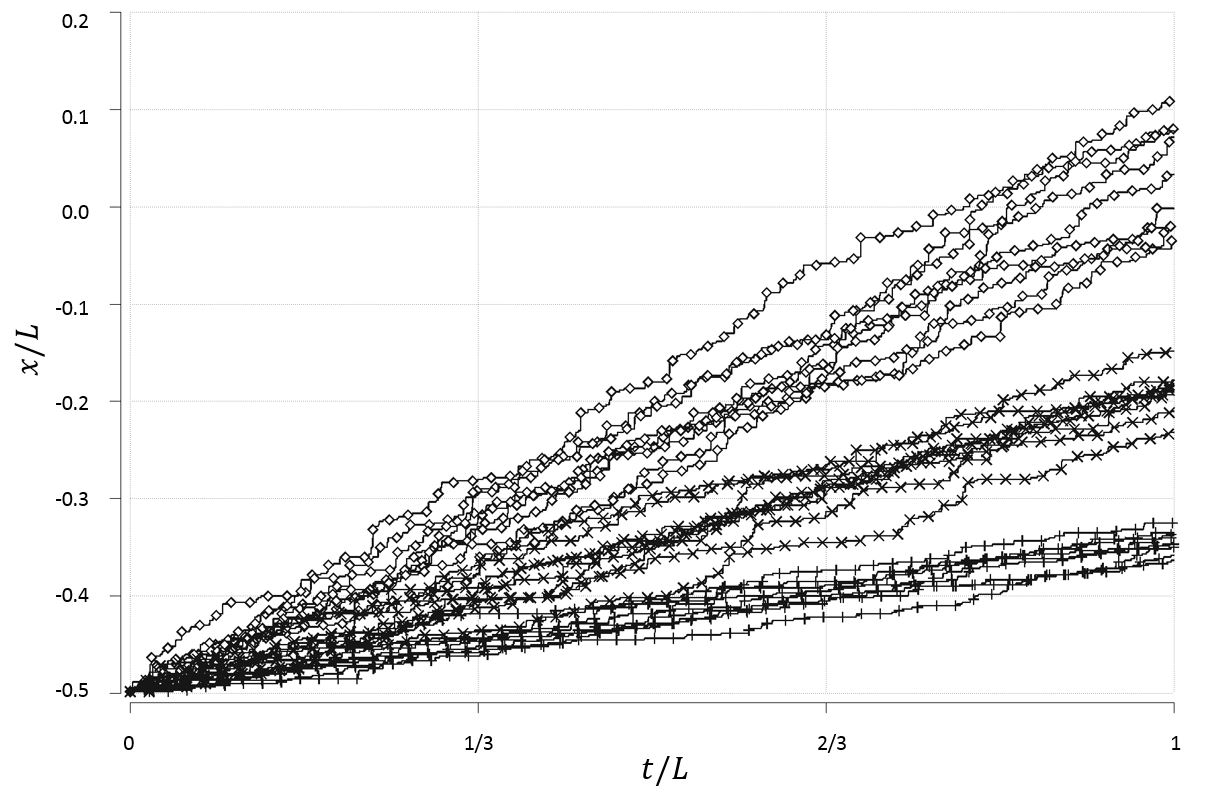}

\captionW{Stochastic trajectories for massive particles with momentum $p=30\pi/L$ and three
different masses, $a\mu=0.25\,(\diamond)$, $a\mu=0.5\,(\times)$ and $a\mu=1\,(+)$; $N=600$.
The position on the vertical axis and the time on the horizontal axis are in units of $L$. \label{MassTraj}}
\end{center}
\end{figure}

From the results shown in Fig.~\ref{H_Kernel}, one could infer not only that particles with larger
mass move more slowly than light particles, but also that their trajectories show less scatter.
This is confirmed by the results shown in Fig.~\ref{MassTraj}. This plot shows trajectories for particles
with the same momentum, $p=30\pi/L$, but different mass, evaluated on a lattice with 600 sites. 
From these results it is not clear, however, to what extent the stochastic nature of the trajectories
will remain visible in the continuum limit, for $a\ra 0$ and $\delta t \ra 0$. To explore this further,
we will compute the effective, macroscopic, velocity of a particle, along with its
fluctuations (i.e., standard deviation) on the average effective velocity.

To arrive at a macroscopic value for the velocity of a particle, we shall first compute its 
displacement after a large number of time steps. This  displacement can be written as
\be
   \Delta x = \sum_{i=1}^M \delta x_i,
\ee
where $M$ is the number of time steps, such that $\Delta t = M\delta t$ is a macroscopic, measurable
time interval. The $\delta x_i$ indicate the individual
particle jumps, which were previously denoted  by $y-x$. The effective particle velocity then follows as
\be
   v_{eff} = \Delta x /\Delta t.
\ee
Note that $\av{\Delta x} = M\av{\delta x}$ and hence $\av{v_{eff}} = \av{v}$.
To find the variance of the average effective velocity 
we first compute the variance of the displacement $\Delta x$,
as
\be
   \av{(\Delta x)^2} - \av{\Delta x}^2 = 
     \av{\sum_{i,j=1}^M \delta x_i \delta x_j } - (\sum_{i=1}^M\av{\delta x_i})^2.
\ee
Using the fact that individual jumps are uncorrelated, this can be written as
\be
   \av{(\Delta x)^2} - \av{\Delta x}^2 = M(\av{(\delta x)^2} -\av{\delta x}^2). \label{var2}
\ee

Like the average displacement $\av{\delta x} = v\,\delta t$, the  average of its square 
is manifestly proportional to the time step size, and hence can be written as
\be
  \av{(\delta x)^2} = \Lambda \delta t.  \label{Lambda}
\ee
  With our choice of units, the 
factor $\Lambda$ has the dimension of a length (or inverse mass); the velocity $v$ 
is dimensionless. The simple 1-D system has only three length scales: 
$a$, $1/\mu$ and $L$. In the non-relativistic limit a similar computation as the one 
leading to the result (\ref{NonRelLimit}) shows that $\Lambda \propto a$.
For relativistic particles, with finite mass, the results shown in 
Fig.~\ref{H_Kernel} suggests that $\Lambda \propto 1/\mu$; for massless particles 
(to be discussed further below), we expect $\Lambda \propto L$.
The parameterization (\ref{Lambda}) implies that Eq.~(\ref{var2}) can be written as
\be
   \av{(\Delta x)^2} - \av{\Delta x}^2 = \Delta t \Lambda(1 - v^2\delta t/\Lambda).
\ee
The time discretization will be such that $\delta t/\Lambda \ra 0$, which implies that the
standard deviation of the effective velocity can be written as 
\be
   \delta v_{eff} = (\av{v_{eff}^2} - \av{v_{eff}}^2)^{1/2} \approx (\Lambda/\Delta t)^{1/2}.
                                                                \label{PathScatter}
\ee
Clearly, the scatter vanishes for non-relativistic particles, for which 
$\Lambda \propto a$ and $a/\Delta t \ra 0$ for $a \ra 0$. 
For massive particles, with $\Lambda \propto 1/\mu$
the stochastic nature of the trajectories will be suppressed when
time and distance scales are long compared to the Compton wave length $1/\mu$ of the particle. 
However, at sufficiently small time or length scales, 
the stochastic nature of the particle will remain to be visible, even in the continuum limit.

The reduction of scatter in the particle trajectories for large mass is illustrated in 
Fig.~\ref{MassTraj}. For three values of the mass parameter, $a\mu = 0.25$, $0.5$ and $1$, 
the figure shows nine typical trajectories of a particles with the same momentum 
$p=30\pi/L$, all starting at the same location, $x/L=-0.5$ on a lattice with $N=600$. 
For comparison, we can compute the classical,
non-relativistic velocity of these particles as well. Using $v = p/\mu = 30\pi/N a\mu$ we
find that these non-relativistic velocities would be $0.63$, $0.31$ and $0.16$ respectively;
these values are fairly close to the average velocities estimated from the trajectories
in Fig.~\ref{MassTraj}. These values are (approximately) $0.52$, $0.29$ and $0.15$ respectively,
which are slightly smaller than the non-relativistic values (as they should be).

The dimensional analysis above suggests that for massless particles the variance will 
be proportional to the only remaining length scale: the system size $L$. This would 
imply that photon trajectories, at least those associated with plane waves, will be intrinsically 
stochastic - even in the continuum limit. 
This will be further explored in the next section.

\subsection{Trajectories for Massless Particles} \label{SubSect3dot3}
Using Eq.~(\ref{TransOneP}), we can write the transition probability 
for a massless free-moving particle to jump from location $x$ to $y$,
with $y \ne x$, in a time interval $\delta t = \xi a$, as
\be
  aT^{(1)}_{y,x}\delta t = {\rm max}\left(0, -{\rm Im}\big( 4\xi a\sum_p |\sin(ap/2)| 
                                                e^{ip(y-x)}\psi(x)/\psi(y)\big) \right).
               \label{TransZeroM}
\ee
Here we used $2-2\cos(ap) = 4\sin^2(ap/2)$. These transition probabilities can also be written as
\be
  aT^{(1)}_{y, x}\delta t = {\rm max}\left(0,-{\rm Im}(\xi \psi(y)/\psi(x))K((y-x)/a) \right).
\ee
with
\bea
   K(n) & = & 4a\sum_p |\sin(ap/2)|e^{iapn}  \\
        & = & (8/N)\sum_{k=1}^{N/2-1}\sin(\pi k/N) \cos(2\pi k n/N) + (4/N)\cos(n\pi),
\eea
where we used the notation $n = (y-x)/a$.
This sum can in fact be evaluated in closed form, by writing the sine and cosine in terms of 
exponentials and summing the resulting geometrical series. This gives the rather simple
result
\be
  K(n) =  (4/N)\sin(\pi/N)/\big(\cos(2\pi n/N) - \cos(\pi/N)\big), \label{Kn}
\ee
from which the transition probabilities follow as
\bea
    aT^{(1)}_{y, x}\delta t & = &  {\rm max}\big(0, {\rm Im} \overline{T}_{y,x} \big)\delta t/L \\
	\overline{T}_{y,x} & = & 	(\psi(y)/\psi(x))
		  4\sin(a\pi/L)/\big( \cos(a\pi/L) - \cos(2\pi(y-x)/L) \big).
				  \label{ZeroMassTrXY}
\eea

\begin{figure}[tbh] 
\begin{center}
\includegraphics[scale=0.42]{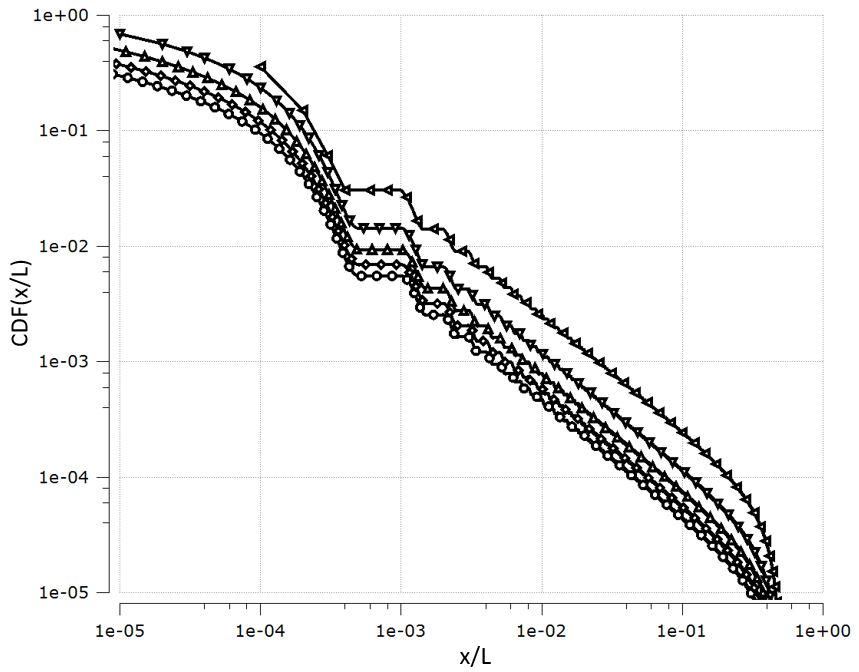}

\captionW{Cumulative probability for a zero-mass particle to jump a distance $x$ or more. The different
curves are for $N=10^4 (\triangleleft)$, $N=10^5 (\bigtriangledown)$, $N=10^6 (\bigtriangleup)$,
$N=10^7 (\Diamond)$ and $N=10^8 (\bigcirc)$ respectively. 
The position on the vertical axis is in units of $L$. \label{ZeroMassCDFx}}
\end{center}
\end{figure}

\begin{figure}[bht] 
\begin{center}
\includegraphics[scale=0.35]{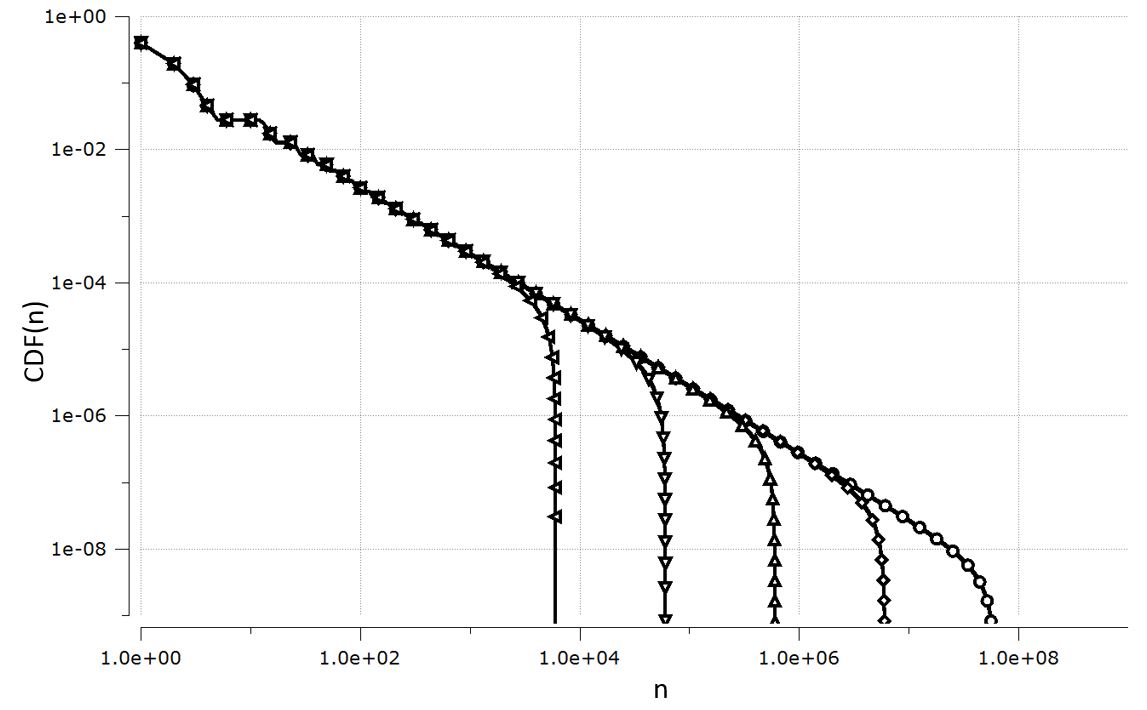}

\captionW{Cumulative probability for a zero-mass particle to jump a distance $n$ or more. The different
curves are for $N=10^4 (\triangleleft)$, $N=10^5 (\bigtriangledown)$, $N=10^6 (\bigtriangleup)$,
$N=10^7 (\Diamond)$ and $N=10^8 (\bigcirc)$ respectively. 
The position on the vertical axis is in units of $a$. \label{ZMCDFn}}
\end{center}
\end{figure}

As anticipated, these transition probabilities do not decrease exponentially with $|y-x|$, but
rather fall off as a power. The behavior of the transition probabilities for large displacements
can be inferred from the curves shown in Figs.~\ref{ZeroMassCDFx} and ~\ref{ZMCDFn}.
The first figure shows the $x$-dependence of the cumulative probability for a particle to jump a distance
$x$ or larger, $CDF(x) = c\sum_{y=x}^{L/2}aT^{(1)}_{y, 0}\delta t$. The constant $c$ is such that
the cumulative distribution is normalized to 1: $CDF(0)=1$. The individual curves are computed
using a fixed value for the wave number $k_0=1000$ and increasing values of $N$, such that the
lattice distance decreases $\propto 1/N$ while the wave length in units of $L$ stays fixed.
These results show that the cumulative jump probability decreases $\propto 1/x$ for intermediate
size jumps ($0.001 \lesssim x/L \lesssim 0.1$). Finite size effects set in for $x/L \gtrsim 0.1$.
This behavior implies that the transition rates decrease $\propto 1/|y-x|^2$ for large jumps distances.
Discretization effects are seen to decrease with increasing $N$, but are still visible, even at $N=10^8$.

The second figure \ref{ZMCDFn} exposes finite size effects on the transition probabilities. 
Now the curves show the $n$-dependence of the cumulative probability 
for a particle to jump a distance $n=|y-x|/a$ or larger, $CDF(n) = c\sum_{i=n}^{N/2}aT^{(1)}_{ia, 0}\delta t$. 
Again, the constant $c$ is such that
the cumulative distribution is normalized to 1: $CDF(0)=1$. The individual curves are computed
using wave numbers that increase with $N$, $k_0=N/12$, such that the wave length in units of the lattice
distance stays fixed, while the system size $L$ increases with $N$. This again shows that
the cumulative jump probabilities decrease as $1/n$ for increasing jump size $n$.

\begin{figure}[tbh] 
\begin{center}
\includegraphics[scale=0.43]{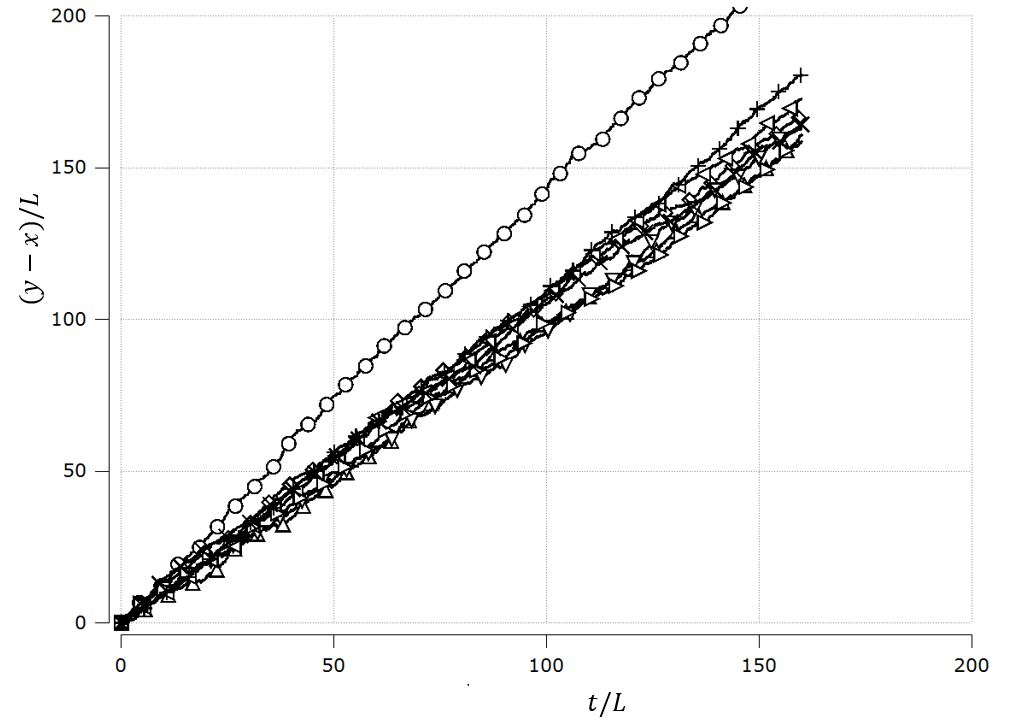}

\captionW{Stochastic trajectories for zero mass particles with different momenta $p=2k\pi/L$
($k = 1 (\circ), 5 (+), 9 (\times), 13 (\diamond), 17 (\bigtriangleup), 21 (\bigtriangledown),
25 (\triangleleft)$ and $29 (\triangleright)$.
The position on the vertical axis and the time on the horizontal axis are in units of $L$ and $N = 4\; 10^5$. \label{ZeroMassTraj}}
\end{center}
\end{figure}

\begin{figure}[htb] 
\begin{center}
\includegraphics[scale=0.55]{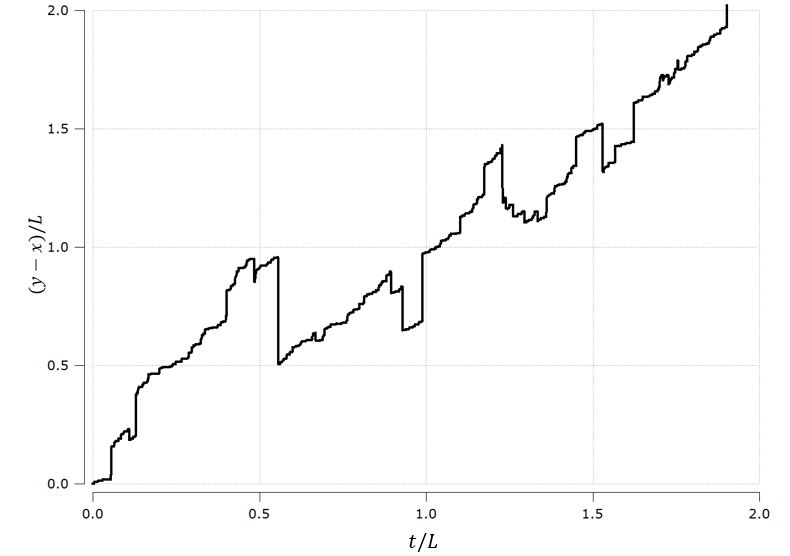}

\captionW{Detailed view of one of the stochastic trajectories shown in Fig.~\ref{ZeroMassTraj}.
Now the total displacement is only  $2L$, which shows that the magnitude of the stochastic jumps can be a
sizable fraction of the system size. \label{ZeroMassTrajDetail} }
\end{center}
\end{figure}

In the previous section, we argued that the variance of the particle velocity would be $\propto L$ for
massless particles. This is confirmed by the scaling behavior shown in Figs.~\ref{ZeroMassCDFx} and ~\ref{ZMCDFn}.
To further confirm this result, we shall compute the average velocity and its variance using
the transition probabilities of Eq.~(\ref{ZeroMassTrXY}).

To allow for a partial analytical treatment, we shall again use the simple fixed momentum
plane wave function of Eq.~(\ref{psix}). The average velocity can then be computed by
first evaluating the average displacement $\delta x = y - x$,
\be
  \av{\delta x} = \sum_x T^{(1)}_{x+\delta x, x}\delta t \delta x  = \xi \sum_n{\rm max}\big(0, \sin(2\pi k_0 n/N) K(n) \big) na,
     \label{avg_dx}
\ee
where we used $p_0 = 2\pi k_0/N$. Next we note that $K(-n)=K(n)$ and since the summation over $n$ runs
symmetrically from $-N/2+1$ to $N/2-1$ (the wave function term is zero for $n=N/2$), it can
also be written as
\be
  \av{\delta x} = \delta t \sum_{n=1}^{N/2-1} \sin(2\pi k_0 n/N) (4/N)\sin(\pi/N)/\big(\cos(2\pi n/N) - \cos(\pi/N)\big) n.
            \label{avgV}
\ee
Writing $\av{v}=\av{\delta x}/\delta t$, it follows that for $N \ra \infty$ at fixed $L$ 
the average velocity can be approximated by an integral, as
\be
  \av{v} \approx \int_0^{L/2} dx \,(x/L)\sin(2\pi k_0 x/L) (4\pi/L)/(\cos(2\pi x/L) - 1),
\ee
or
\be
  \av{v} \approx (1/\pi)\int_0^\pi dt\, t \sin(k_0 t)/(\cos(t) - 1).
\ee
Finally, in the limit $k_0 \ra \infty$, the substitution $k_0 t = z$ shows that the integral can be evaluated as 
\bea
 \av{v} & \approx & (1/\pi)\int_0^{k_0 \pi} dz\, z \sin(z)/k_0^2(\cos(z/k_0) - 1) \\
        & \approx & (2/\pi)\int_0^{\infty} dz\, \sin(z)/z = 1.
\eea

To arrive at this result, both the continuum limit $a \ra 0$ (or equivalently, $N\ra \infty$) and the
small wave length limit $k_0 \ra \infty$ were taken. To see how much the average particle
velocities deviate from 1 for finite values of $N$ and $k_0$, the summation (\ref{avgV}) can be 
done numerically. This leads to the results in table \ref{tableOne}, which show that
the impact of $N$ is fairly small (at least once $N \gtrapprox 10^5$).

Of course it is gratifying to find that the average velocity of massless 
particles is $\approx 1$. However, this does not imply that the individual trajectories
resemble the paths of a classical particle moving at the velocity of light. 
This would require that the variance of particle displacement $\delta x$ either 
vanishes for $a \ra 0$ or has a sufficiently small finite value. The discussion 
of the previous section already indicated that this will not be
the case when the particle's wave function is a plane wave. Then the only available
length scale is the system size and the displacement variance will be $\propto L\delta t$. 
This can now be explicitly verified using the exact result for the Hamiltonian kernel. 
A slight modification of Eq.~(\ref{avg_dx}) gives
\be
 \av{(\delta x)^2} = \sum_x T^{(1)}_{x+\delta x, x}\delta t (\delta x)^2  
            = L\delta t \sum_n{\rm max}\big(0, \sin(2\pi k_0 n/N) K(n) \big) (n/N)^2.
			 \label{dx2}
\ee
Now it is unfortunately not possible to use a sign flip of $n$ to remove 
the ``max'' prescription and the summation has to be performed numerically. 
The results of numerically computing $\av{(\delta x)^2}/L\delta t$ from Eq.~(\ref{dx2})
have been included in table \ref{tableOne}. 
This shows that the variance of $\Delta x$ converges to $\approx 0.281 L\delta t$.
Therefore, the length scale $\Lambda$ introduced in Eq.~(\ref{Lambda}) approximately 
equals $0.281 L$ for massless bosons in 1-D that move according to a non-localized
plane wave.

\begin{table}
\begin{center}
\begin{tabular}[tbh]{|r||c|c|} \hline
\multicolumn{3}{|l|}{\bfseries $k_0 = N^{1/2}$, $L$ fixed ($a=L/N$)} \\ \hline
$N$ & $\av{\delta x}/\delta t$ & $\av{(\delta x)^2}/L\delta t$ \\
\hline \hline
$10^3$ & $0.6955$     & $0.2868$  \\ \hline
$10^4$ & $0.9030$     & $0.2734$  \\ \hline
$10^5$ & $1.0113$     & $0.2810$  \\ \hline
$10^6$ & $0.9945$     & $0.2808$  \\ \hline
$10^7$ & $0.9995$     & $0.2809$  \\ \hline
$10^8$ & $0.9998$     & $0.2809$  \\ \hline
\end{tabular}

\captionW{Numerically computed results for the average velocity and its variance for different values of $N$.
The wave number increases as $k_0 = N^{1/2}$ to ensure that the wave length is sufficiently small while
the lattice distance decreases.  \label{tableOne}}
\end{center}
\end{table}

In Fig.~\ref{ZeroMassTraj} we show trajectories of a zero mass particle with 
eight momenta ranging from $p=2\pi/L$ to $p=58\pi/L$. This confirms that the effective
particle velocity is $\approx 1$ for large momenta. Only for the two lowest momenta,
$p=2\pi/L$ and $p=10\pi/L$, the velocity is significantly larger than 1. For larger momenta,
$p \gtrsim 20\pi/L$, the $p$-dependence is less than the scatter between individual
trajectories.
These high-momentum particles have remarkably straight trajectories
with a slope roughly equal to 1 (the velocity of light). 
Note, however, the total travel time is quite long, 
such that the particles move $\approx 150$ times around the universe. On that scale, 
the scatter in displacements becomes relatively small (in accordance with the results 
for the standard deviation shown in Eq.~(\ref{PathScatter})). The strong stochasticity
is, however, clearly visible if the trajectories are viewed on the smaller scale of just a few L.
This can be seen in Fig.~\ref{ZeroMassTrajDetail}.

\section{Two-particle State without Interaction} \label{Sect4}
Before turning to particles with interaction, it is interesting to look at a multi-particle
state in the free theory. This will allow illustrating some (well-known) peculiarities
of Bohm-type particle trajectories.
The simplest multi-particle state is a state with just two particles. Since the interaction
term is still absent, this implies that the dynamics of this state will be fully governed by
the free Hamiltonian in the 2-particle sector,
\bea
   \bra{y_1 y_2}\hat{H_0}\ket{x_1 x_2} & = &
   \sum_p \omega_p\bra{0}\ha_{y_1}\ha_{y_2}(\had_p \ha_p + \half L)\had_{x_1}
   \had_{x_2}\ket{0} \\
   & = & \hh{y_1}{x_1}\delta_{y_2,x_2} + \hh{y_1}{x_2}\delta_{y_2,x_1} \\
   & + & \hh{y_2}{x_1}\delta_{y_1,x_2} + \hh{y_2}{x_2}\delta_{y_1,x_1}. \label{Hyyxx}
\eea
 The Schr\"odinger equation in the 2-particle sector looks like
\be
  i\dt \psi(x_1,x_2)  =   \sum_{y_1,y_2} \bra{x_1 x_2}\hat{H}\ket{y_1 y_2}\psi(y_1,y_2),
				\label{SchrEq2Pfree}
\ee
where the state vectors and Hamiltonian matrix now are defined on a  $N(N+1)/2$ dimensional space.
Eq.~(\ref{Hyyxx}) shows that the dynamics is still essentially driven by the 1-particle Hamiltonian
matrix elements (\ref{HxyOneP}).

Using these 2-particle Hamiltonian matrix elements, it is straightforward to also compute
the transition rates $T^{(2)}_{y_1 y_2, x_1 x_2}$, which can then be used to generate
particle trajectories. As before, we shall use an initial state with fixed momenta for the two
particles,
$\ket{\Psi_0} = \had_{p_1}\had_{p_2}\ket{0}$ (with $p_1 < p_2$). 
In the particle configuration representation, this state can
be written as
\be
   \ket{\Psi_0}=\sum_{x_1 \le x_2}\psi^{(2)}(x_1,x_2)\ket{x_1 x_2},
\ee
with $\ket{x_1 x_2} = \had_{x_1}\had_{x_2}\ket{0}$ and
\be
   \psi^{(2)}(x_1,x_2) = (2L^2)^{-1/2}\big( e^{i(x_1 p_1 + x_2 p_2)} + e^{i(x_1 p_2 + x_2 p_1)}\big).
   \label{psi2}
\ee
Note that this is (as before) a non-localized state, where the position eigenstates are now
highly entangled. In particular, when the two
particles have equal but opposite momenta, $p_2 = -p_1$, this state is a real-valued
eigenstate of the Hamiltonian. Since also the Hamiltonian matrix elements are real, this
implies that all transition rates for a particle to move to a different location are zero. This suggests
that both particles in this state in fact move according to the average momentum in this state -
which is zero if $p_2 = -p_1$. This collective movement according to the average momentum
holds for all states of the form (\ref{psi2}) as can be checked explicitly in the non-relativistic
limit where the dynamics reduces to the causal Bohm-dynamics.

\begin{figure}[tbh] 
\begin{center}
\includegraphics[scale=0.42]{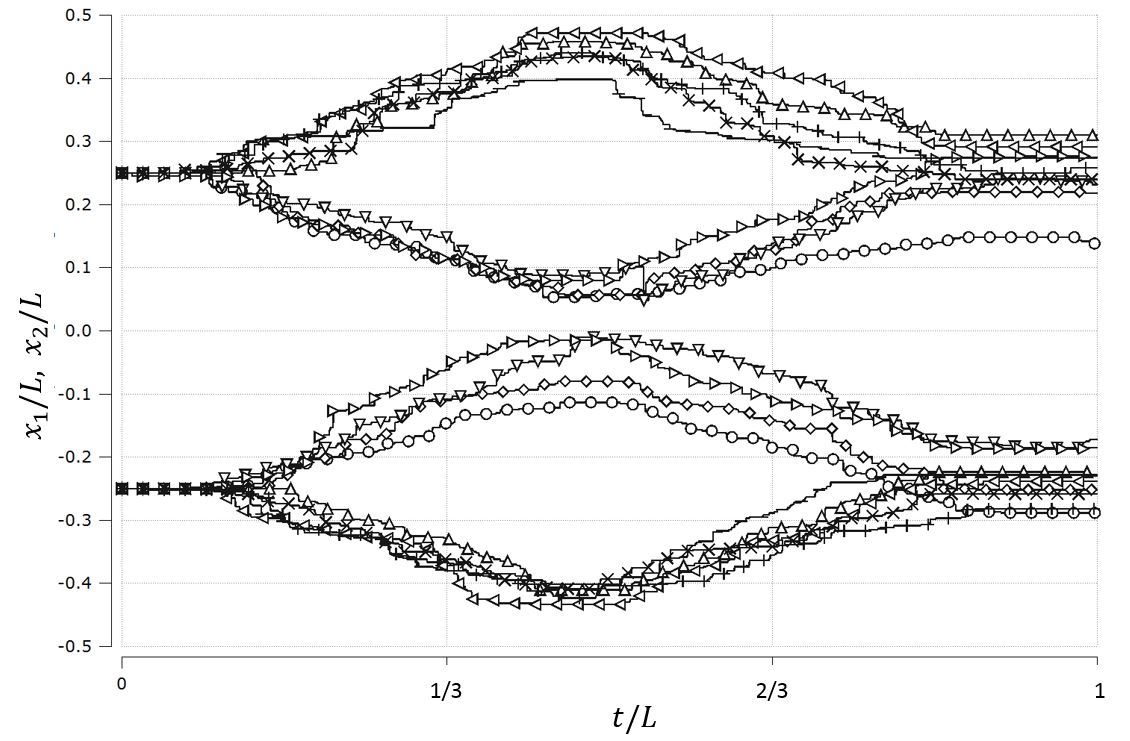}

\captionW{Trajectories for a 2-particle state without interactions. The initial state has
localized Gaussian wave functions with a width of $0.075$ and peak positions at
$x_1^0=0.25$ and $x_2^0=0.75$ respectively.
The initial momenta are $\pm 30\pi/L$, the mass parameter $a\mu = 0.25$ and the periodic 
lattice has 600 sites. The plot show 10 trajectories, 
which are labeled by different symbols. \label{TwoFreeParticles} }
\end{center}
\end{figure}

In order to show more separated trajectories for the two particles, we shall ``decohere'' the two
particles by localizing them with a Gaussian wave packet. An example of such a localized
2-particle wave function is given by
\be
   \psi^{(2)}(x_1,x_2) = 
   c \big( e^{-(x_1 - x^0_1)^2/2\sigma -(x_2 - x^0_2)^2/2\sigma + i(x_1 p_1 + x_2 p_2)}
    + (x_1 \leftrightarrow x_2) \big).
\ee
The constant $c$ is a normalization constant and we use the same packet width $\sigma$ for
both particles. This state represents an initial condition in which one particle is localized
near $x^0_1$, the other near $x^0_2$; one of the particles has momentum $p_1$,
the other has momentum $p_2$. Note that these particles of course cannot be
identified, hence it is not clear which particle will have momentum $p_1$ or $p_2$.

The stochastic BBB trajectories for such an evolving state are shown in Fig.~\ref{TwoFreeParticles}.
The initial position for the two particles was chosen the same for all 10 trajectories at
$x_1^0/L=0.25$ and $x_2^0/L = 0.75$, the initial package width is $0.075L$ and the two particles
have an equal but opposite momentum $p=30 \pi/L$. As may have been expected, each particle
can move in either of the two directions, but their effective velocities will always be opposite.
It is also worth noting that the particle trajectories do not cross, but ``bounce back''. I.e.,
the stochastic dynamics favors transitions where the two particles swap their location, as soon
as the guiding wave packages are  sufficiently overlapping to make such transitions likely.

\section{Particle Creation and Annihilation} \label{Sect5}
In order to explore how particle creation and annihilation is represented by the BBB
trajectories, we not only have to extend the state space to include extra particles,
but also have to add the cubic interaction term (\ref{Hint}) to the
Hamiltonian. Now we have to numerically solve the coupled multi-particle wave 
function evolution as well as the associated particle configuration dynamics. 
Solving the  Schr\"odinger
equation  (\ref{SchrEq}) numerically quickly becomes very challenging if the number of 
degrees of freedom, i.e., the number of lattice sites and particles, increases. 
With limited compute resources we are forced to use a highly simplified system
- without bothering too much about the impact on the physical content of the simplified theory.
The main purpose here will be an illustration of how BBB dynamics captures particle
creation and annihilation in an interacting QFT.

Since we will be using the particle configuration representation,
 we can easily choose to restrict the dynamics to processes involving a limited 
number of particles. This can be achieved by using a truncated Hilbert which is a 
Fock-space with a maximum number of particles; Here we shall choose
the very low value $M_{max} = 2$ in Eq.~(\ref{unity}) and assume that this will still capture enough
of the relevant dynamics to illustrate particle creation and annihilation.
The cubic self-interaction term in this basis can then only
change the particle content by $\pm 1$. If the initial state has 1 particle,
we obtain a self-consistent dynamics where the interaction Hamiltonian  supplies
matrix elements between 1- and 2-particle states, as will be shown in
more detail below. 

The truncated Fock space of this system is spanned by the states
\be
  {\ket{x_1}}_{x_1\in L_N},\;{\ket{x_1 x_2}}_{x_1,x_2 \in L_N; x_1\le x_2}.
\ee
This is a $ N + \half N(N+1)$ dimensional space, if the 1-D lattice has $N$ sites.
We use the notation $L_N = {-L/2+a,\cdots,L/2}$ for the set of $N$ lattice
locations. An arbitrary state can then be written as
\be
 \ket{\Psi} =  \sum_{x_1\in L_N} \psi(x_1) \ket{x_1} 
         + \sum_{x_1,x_2 \in L_N; x_1\le x_2}\psi(x_1,x_2)\ket{x_1 x_2}. \label{PsiTrunc}
\ee
Recall that we choose to exclude a vacuum state contribution in the (initial) state.
To specify the Hamiltonian on this space, the following non-zero matrix elements have
to be computed:
\be
 \bra{y_1}\hat{H}\ket{x_1},\;\bra{y_1}\hat{H}\ket{x_1 x_2},\;
 \bra{y_1 y_2}\hat{H}\ket{x_1},\;
 \bra{y_1 y_2}\hat{H}\ket{x_1 x_2}. \label{H_elements}
\ee  
The Schr\"odinger equation for states of the form (\ref{PsiTrunc})
takes the form,
\bea
  i\dt \psi(x_1)         & = & \sum_{y_1} \bra{x_1}\hat{H}\ket{y_1}\psi(y_1) 
                          + \sum_{y_1,y_2} \bra{x_1}\hat{H}\ket{y_1 y_2}\psi(y_1,y_2), \\
  i\dt \psi(x_1,x_2) & = & \sum_{y_1} \bra{x_1 x_2}\hat{H}\ket{y_1}\psi(y_1) 
                + \sum_{y_1,y_2} \bra{x_1 x_2}\hat{H}\ket{y_1 y_2}\psi(y_1,y_2).
				\label{SchrEq2P}
\eea
This coupled set of differential equations can be discretized and solved numerically once the
four sets of matrix elements (\ref{H_elements}) have been (pre)computed. The first set of matrix
elements, $\bra{y_1}\hat{H}\ket{x_1}$ were used already to compute 1-particle trajectories in
the sections above. There are $N \times N$ of these elements, which can be computed using Eq.~(\ref{HxyOneP}).
There are $N(N+1)/2 \times N(N+1)/2$ matrix elements $\bra{y_1 y_2}\hat{H}\ket{x_1 x_2}$ 
that connect two 2-particle states. These matrix elements also 
feature in the 2-particle model without interaction and were discussed already in section \ref{Sect4}.

\begin{figure}[th] 
\begin{center}
\includegraphics[scale=0.37]{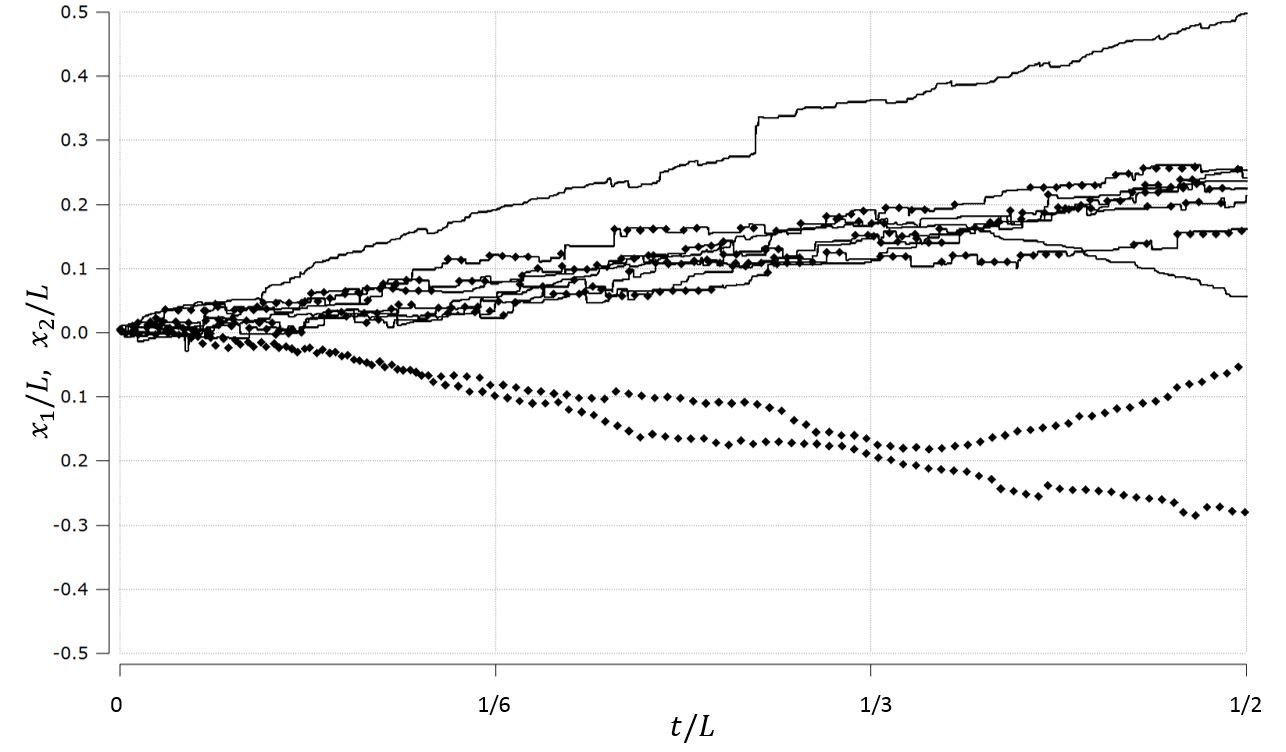}

\captionW{Ten trajectories for scalar bosons with self-interaction strength $a^2\lambda = 0.5$ and mass parameter $a\mu = 0.25$ on
a lattice with $N=600$ sites.
The initial state is a 1-particle state with momentum  $p=\pm 30\pi/L$, and initial position $x/L=0$ for all trajectories.
The solid diamonds indicate the presence of a second particle. Note that for all trajectories except two, this second
particle propagates close to the first. \label{MultiP} }
\end{center}
\end{figure}

\begin{figure}[ht] 
\begin{center}
\includegraphics[scale=0.37]{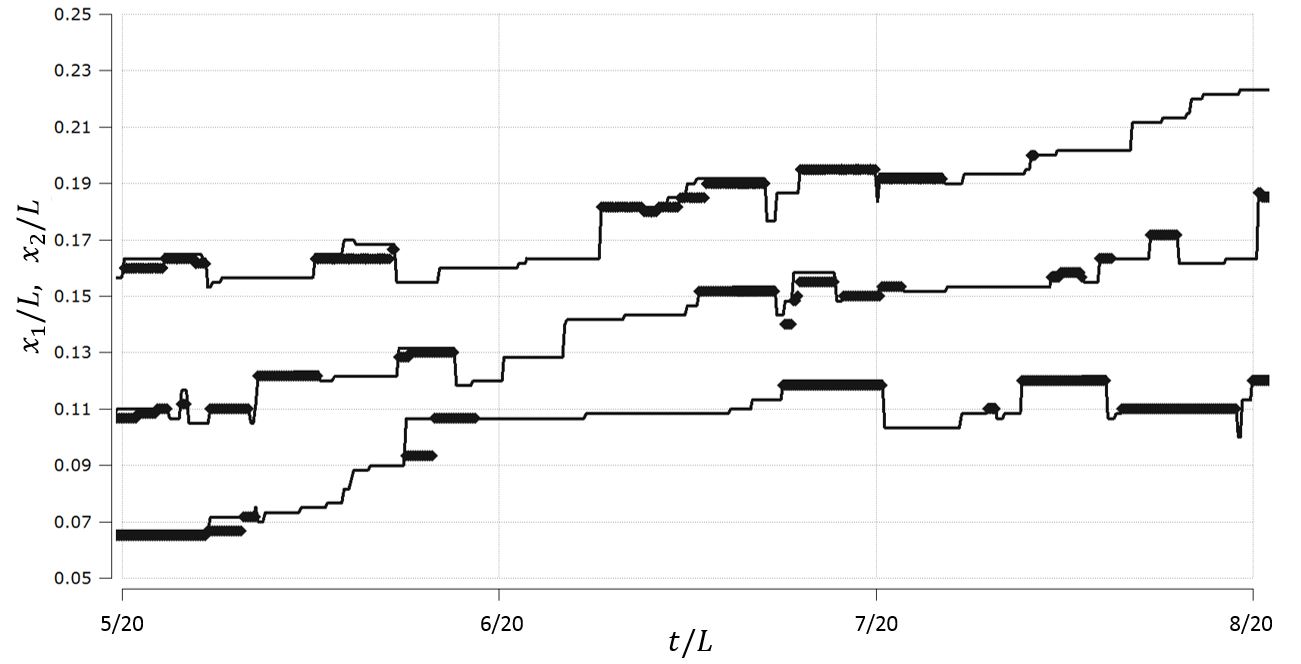}

\captionW{Detailed view on three trajectories from Fig.~\ref{MultiP}. The fat lines (which are strings of diamond symbols)
represent the presence of a second particle. \label{MultiPzoom} }
\end{center}
\end{figure}

In order to accommodate  particle creation and annihilation, we have to compute the matrix elements
that connect 2-particle states to 1-particle states:
\be
    \bra{y_1}\hat{H}\ket{x_1 x_2}  = 
       \lambda \sum_x \sum_{p_1,p_2,p_3} (8\omega_{p_1}\omega_{p_2}\omega_{p_3})^{-1/2} e^{i(p_1 - p_2 - p_3)x}
	           \bra{0}\ha_{y_1}\had_{p_1} \ha_{p_2}\ha_{p_3}\had_{x_1}\had_{x_2}\ket{0}.
\ee
This can be evaluated as
\be
  \bra{y_1}\hat{H}\ket{x_1 x_2}  = 2\lambda \sum_x S(y_1 - x) S(x_1 - x) S(x_2 - x), 
     \label{Hyxx}
\ee
with
\be
 S(x) = \sum_p (2 \omega_p)^{-1/2} e^{ipx}.
\ee
There are $N \times N(N+1)/2$ of these matrix elements, which have a locality that is now governed by the two-point
function $S(x)$.
Note that we have adopted the normal ordering prescription for the operators in the interaction term.
Since the matrix elements are real-valued, the matrix elements that connect 1-particle states to
2-particle states follow easily as $\bra{y_1 y_2}\hat{H}\ket{x_1} = \bra{x_1}\hat{H}\ket{y_1 y_2}$.

Since the state space is now significantly larger 
and since we have to include the computationally heavy interaction term (\ref{Hyxx}) 
in the Hamiltonian, the numerical effort of solving Eq.~(\ref{SchrEq2P}) 
and the associated transition rates (\ref{trans}), is much larger than for the non-interacting systems. 
As a practical point (which further increases the computational burden),
we note that the time discretization of (\ref{SchrEq2P}) should be done such, that the state 
evolution is unitary  to a high accuracy (i.e., the numerical time evolution must accurately preserve 
the norm of the state vector).
The actual computation of the stochastic particle location evolution is, once the wave function has been
computed, relatively less time consuming. Hence, it is advantageous to compute a large number of trajectories
concurrently.

For the illustration of the BBB trajectories of interacting particles we shall again
use a lattice with $N=600$ sites, a mass parameter $a\mu=0.25$ and the same initial states as used for the free theory.
I.e., we use
\be
   \psi_0(x_1) = L^{-1/2}e^{ip_0 x_1},\; \psi_0(x_1,x_2) = 0, \label{ini_psi}
\ee
with $p_0 = 30\pi/L$. From the trajectories shown in Fig.~\ref{MassTraj} we infer
that, without self-interaction, a particle with the same mass and with this initial momentum,
has trajectories that are recognizably straight lines, with an average velocity approximately
equal to $0.52$.
With self-interaction, we expect that an extra particle can be created along the way, and if this sufficiently
would resemble a ``cloud of virtual particles'' the resulting trajectories should be that of a particle with
an increased mass, and hence a reduced velocity. 

This behavior is indeed observed in the numerical simulations. We find that spontaneous particle creation happens
when the self-interaction is sufficiently strong. For values $a^2\lambda <0.1$ the simulated time period is too short to see 
particle creation. For $a^2\lambda \gtrapprox 0.1$ this changes, and a significant number of particle creation
events can be observed in a typical simulation.
This particle creation is illustrated by the result shown in Fig.~\ref{MultiP}, which was obtained using $a^2\lambda = 0.5$.
This figure shows ten trajectories
that all start from $x/L=0.5$. The full lines represent the paths of a particle, which is often  accompanied by a second
particle, indicated by the strings of full diamonds. For most trajectories, this second particle moves along at a  close
spacing from the first particle (often occupying the same lattice sites). This can be seen more clearly
in Fig.~\ref{MultiPzoom}, which is a zoom-in on some of the trajectories from Fig.~\ref{MultiP}. 

Even though the trajectories
show significant scatter, we can read off an average velocity for the paired particles of around $0.4$, which
is clearly lower than the velocity $0.52$ of the particle without self-interaction (cf. Fig.~\ref{MassTraj}).
Two trajectories show a deviant behavior: here a second particle almost immediately splits up
from the first and travels in opposite direction.
Since this simplified model does not support the creation of additional particles, these isolated particles propagate
as ``undressed'' particles, with a larger velocity than the dressed particles. Rather surprisingly, in one of these
split pairs, the particles reverse their directions, at $t/L \approx 1/3$, and start approaching each other. This may
be an artifact of the simplifications in this truncated 2-particle model.

\section{Summary and Discussion} \label{Sect6}
In this paper we elaborated on Bell's proposal \cite{Bell} for computing
particle trajectories for Quantum Field Theory (QFT). In order to make this work
as rigorous and explicit as possible, we used a simple QFT which describes interacting bosons on a
1-dimensional (spatial) lattice. We reviewed how the full quantum dynamics, as it is provided by the
Schr\"odinger equation can be either expressed in a ``field representation'', where the
wave function is a defined on the space of all (lattice) field configurations; or it can be
expressed in a ``particle configuration representation'', where the wave function is 
defined on a (possibly truncated) Fock space spanned by states in which particles have
specified locations. This Fock space is unitary equivalent to the more commonly used Fock
space in which basis states describe particles with specified momenta.
Bell's approach can then be applied to define transition rates between particle configurations;
i.e., using his approach, we can provide probabilities for a specific particle configuration 
at time $t$ to jump to another configuration at time $t + \delta t$. These particle configurations
are represented as integer valued eigenvalues of a configuration operator, $\hat{C}$, defined
in Eq.~(\ref{config}).

We showed that, as expected, the stochastic trajectories of the particles reduce to the causal trajectories defined
by Bohm \cite{Bohm52,BohmHiley} in the non-relativistic limit. We furthermore argued, and illustrated
this with numerical examples, that the stochastic nature of the particle trajectories is suppressed
when the displacement extends over macroscopic distances (i.e., distances much larger than the particle's
Compton wave length). Also massless bosons are found to have well-defined trajectories, which have the gratifying
property that -- irrespective of their momentum -- the average velocity is equal to the speed of light (in the continuum
limit). However, when these trajectories are computed from a fully delocalized (plane wave) particle state, 
the stochastic nature of the trajectories will only be suppressed at
scales much larger than the system size (i.e., the size of the universe). This length scale
will presumably be replaced by the size of the wave packet or coherent state, when the massless
particles would be guided to propagate within the bounds of a localized wave function.

Finally we have shown that particle creation and annihilation is naturally accommodated by the BBB particle configuration
trajectories. We used a 2-particle version of the model with a simple cubic self-interaction
to illustrate that spontaneous particle creation happens once the coupling strength is sufficiently
large. This then leads to an effective particle propagation at a lower velocity than achieved in
the free theory. I.e., we tentatively recover the result that a dressed particle has a larger mass
than the bare particle of the free theory.

With this work, we hope to have demonstrated that particle trajectories can be obtained equally
well from relativistic Quantum Field Theory as from non-relativistic Quantum Mechanics. It is
interesting that the strict locality of the particle propagation, which results in a causal dynamics
in the non-relativistic limit, is not maintained in the relativistic theory. Hence, stochasticity
or randomness appears to be present at a fundamental level - at least if this formulation of QFT is adopted.
Obviously, these stochastic trajectories will not transform in a relativistically covariant manner, which 
reinforces an apparent conflict between Bohm dynamics and Lorentz invariance. 
See \cite{DurrGoldsteinNorsen} for further discussion and references on this issue.

In the BBB formulation, it is clear that a specific foliation of space-time has to be chosen before the stochastic
trajectories can be computed. However, this need not be seen as a problem for the BBB formulation:
It should be quite acceptable that a specific realization of the QFT, i.e., a specific simulation of a
concrete universe, requires specifying a split between space and time. Such a ``spontaneous'' breaking
of the underlying Lorentz invariance of the QFT of course in no way diminishes the value or importance of 
this symmetry. One should furthermore
consider that the specifics of the particle trajectories (be they causal Bohm trajectories in
quantum mechanics, or stochastic BBB trajectories in QFT) typically elude direct detection. Hence,
the specific space-time split realized in our universe is likely bound to remain unobservable in practice.

The stochastic BBB dynamics, by its very nature, also illustrates how the inherent non-locality of quantum mechanics
carries over to QFT: the full particle content of the universe is described by a single configuration
index. Time evolution of the collective particle content of this universe is captured by the time
evolution of this single, integer, configuration index.

In summary, we presume to have demonstrated that the same power of explanation that is achieved
by classical physics theories can also be achieved by relativistic QFT. 
As is the case in its classical counterpart,
a QFT can generate the dynamics for fields, e.g, the electromagnetic field, as well as
the dynamics for the particles that interact with this field. To be more specific, we would
argue that a similar application of the BBB approach described here for a simple 1-D scalar QFT,
can (in principle) be applied to the Standard Model. This assumes that we
can start from a lattice-regularized version of the Standard Model, in which the full,
non-perturbative quantum dynamics can be captured using a configuration Fock space; this 
Fock space is spanned by states generated using the creation operators associated 
with the fundamental fields in the model (cf. Eqs.~(\ref{ketx}-\ref{config})).
In this way, the BBB formulation can be used to provide the dynamics for 
the gauge fields as well as dynamics for the fermionic and bosonic particles in the Standard Model.
For QED this would imply, that we could directly simulate the dynamics of the electromagnetic
field and its interactions with the charged particles - as we can do in classical Maxwell theory. 
However, unlike in simulations of the classical field theory, the dual wave-particle nature of 
the photon would  be naturally exposed in simulations of the quantum field theory.

Of course it is not clear to what extent such numerical simulations will be achievable in practice.
It is already a formidable numerical challenge to solve a 2-particle Schr\"odinger equation for particles
moving in two or three dimensions. However, through further advances
in compute power and clever algorithms, it might well be achievable already now or in the near future,
to expose the microscopic dynamics of (sectors of) the Standard Model through direct simulation. 
It would be quite remarkable, if the BBB formulation could be used, for example, to show in 
a real-time simulation how a quark and anti-quark interact with gluons to form a bound meson state.

\end{document}